\documentclass[
aps,%
12pt,%
final,%
notitlepage,%
oneside,%
onecolumn,%
nobibnotes,%
nofootinbib,%
superscriptaddress,%
noshowpacs,%
centertags]%
{revtex4}
\usepackage{amsmath}
\usepackage{amsthm}
\usepackage{graphics}
\usepackage{epstopdf}
\newtheorem{thh}{Theorem}
\begin{document}

\title{The discrete family symmetries as the possible solution to the flavour problem\\
}

\author{\firstname{Bartosz}~\surname{Dziewit}}
\email{bartosz.dziewit@us.edu.pl}
\affiliation{Institute of Physics, University of Silesia\\
Uniwersytecka 4,  Katowice, Poland
}
\author{\firstname{Jacek}~\surname{Holeczek}}
\email{jacek.holeczek@us.edu.pl}
\affiliation{Institute of Physics, University of Silesia\\
Uniwersytecka 4,  Katowice, Poland
}%

\author{\firstname{Monika}~\surname{Richter}}
\email{mrichter@us.edu.pl}
\affiliation{Institute of Physics, University of Silesia\\
Uniwersytecka 4,  Katowice, Poland
}%

\author{\firstname{Sebastian }~\surname{Zajac}}
\email{s.zajac@uksw.edu.pl}
\affiliation{Faculty of Mathematics and Natural Studies \\
Cardinal  Stefan Wyszynski University in Warsaw\\
Dewajtis 5, Warszawa, Poland
}

\author{\firstname{Marek }~\surname{Zralek}}
\email{marek.zralek@us.edu.pl}
\affiliation{Institute of Physics, University of Silesia\\
Uniwersytecka 4,  Katowice, Poland
}%
%


\begin{abstract}
In order to explain the fermions' masses and mixing parameters appearing in the lepton sector of the Standard Model, one proposes the extension of  its symmetry. A discrete, non-abelian subgroup of $U(3)$ is added to the gauge group $SU(3)_{C}\times SU(2)_{L}\times U(1)_{Y}$. Apart from that, one assumes the existence of one extra Higgs doublet. This article focuses mainly on the mathematical theorems and computational techniques which brought us to the results.
\end{abstract}
\maketitle
\section{Introduction} 
The flavour problem constitutes one of the most serious disadvantages of the Standard Model \cite{fromth}. It doesn't allow us to predict theoretically neither the particles' masses nor the parameters describing $U_{PMNS}$ mixing matrix (Pontecorvo-Maki-Nakagawa-Sakata matrix).
This fact clearly points to some extension of the present-day theory.

In our approach, one adds some new, non-abelian discrete symmetry $G_{F}$ to the  Standard Model's gauge group. This step can be partially justified bearing in mind the enormous success of the \textsl{tribimaximal mixing} \cite{Tbm}  which could be explained by such a group.
Since 2004 many similar ideas have already  been  widely studied, but gave no reasonable results. For this reason, we have decided to go one step further and widen the scalar sector as well. To our knowledge, not so many attempts have been ventured in this direction, so far. Only one article concerning such a model (with 3 Higgs doublets) has been found in the literature \cite{modragon}. This fact motivated us to investigate the model with two Higgs doublets with a lot of possible choices for group $G_{F}$ \cite{dziewit}.

\subsection{\label{sec:level2}The main concept}

To start with, it is necessary to introduce the Yukawa part of the considered  model:
\begin{equation}
\mathcal{L}_{Y}=-\sum_{\alpha,\beta=e,\mu,\tau}\sum_{j=1,2} [(h^{l}_{j})_{\alpha\beta} \bar L_{\alpha L}\Phi_{j}l_{\beta R}+(h^{\nu}_{j})_{\alpha\beta}\bar L_{\alpha L}\tilde \Phi_{j} \nu_{\beta R}],
\end{equation}
where
\begin{itemize}
 
 \item  $e,\mu,\tau$- three different flavors of charged leptons and neutrinos, 
 \item $L_{\alpha L}=\left[\begin{array}{c}\nu_{\alpha L}\\l_{\alpha L}\end{array}\right]$- the lepton doublet composed of left-handed charged lepton and neutrino field of given flavor $\alpha$,
 \item  $l_{\beta R},\nu_{\beta R}$- the right-handed charged lepton and neutrino field of given flavor $\beta$,
 \item $\Phi_{1,2}=\left[\begin{array}{c}\Phi_{1,2}^{+}\\\Phi_{1,2}^{0}\end{array}\right]$ - two Higgs doublets, 
 \item $\tilde \Phi_{1,2}$=$i\sigma_{2}\Phi^{*}_{1,2}$,
 \item $h^{l}_{1,2},h^{\nu}_{1,2}$ - Yukawa couplings. 
\end{itemize}
 We assumed here the minimal extension of the Standard Model: to get the masses for neutrinos, three right-handed neutrino singlets have been added.
 
In principle, four Yukawa couplings $h_{1,2}^{l,\nu}$ occurring in the model are arbitrary 3-dimensional matrices. This lack of any restrictions causes that the masses of fermions and mixing parameters are treated as free  parameters within the Standard Model. 

In order to solve this problem, one can try to constrain the allowed masses by imposing some additional symmetry $G_{F}$ on the Lagrangian.  In simple terms, we expect  Eq.(1) to be invariant  under the  following set of transformations:
\begin{equation}
L'_{\alpha L}=(A^{L})_{\alpha,\beta}L_{\beta L},\quad\quad l'_{\beta R}=(A^{l})_{\beta,\gamma}l_{\gamma R},
\end{equation}
$$
\nu'_{\beta R}=(A^{\nu})_{\beta,\delta}\nu_{\delta R},\quad\quad \Phi_{i}'=(A^{\Phi})_{i,k}\Phi_{k}.
$$

In the above formulas $A^{L},A^{\nu},A^{l}$ stand for \textbf{3}-dimensional irreducible (in general different) representations of $G_{F}$. On the other hand, $A^{\Phi}$ denotes the \textbf{2}-dimensional irreducible representation of the same group. The dimensions of these representations follow from the existence of \textbf{3} families of fermions  and \textbf{2} Higgs doublets which are assumed in the model.

The requirement of the invariance of Eq.(1) under the transformations given in the above-mentioned formulas leads to the equations restricting the form of Yukawa couplings \cite{ludl2}:
\begin{eqnarray}
\sum_{i=1,2} (A^{\phi})_{i,k}(A^{L})^{\dag}_{\alpha\gamma}(h^{l}_{i})_{\gamma\delta}(A^{l})_{\delta\beta}=(h^l_{k})_{\alpha\beta},\nonumber\\
\sum_{i=1,2} (A^{\phi})^{*}_{i,k}(A^{L})^{\dag}_{\alpha\gamma}(h^{\nu}_{i})_{\gamma\delta}(A^{l})_{\delta\beta}=(h^{\nu}_{k})_{\alpha\beta},
\end{eqnarray} 
which after some indices manipulation take the form of 2 eigenproblems to the eigenvalue 1: 
\begin{equation}
\mathbf{N_{1}}\Gamma^{l}=\Gamma^{l},\quad\quad \mathbf{N_{2}}\Gamma^{\nu}=\Gamma^{\nu}, 
\end{equation}
where
\begin{eqnarray}
 (\Gamma^{l,\nu})^{T}=\left[\begin{array}{c}(h_{1}^{l,\nu})_{11},...,(h_{1}^{l,\nu})_{33},(h_{2}^{l,\nu})_{11},...,(h_{2}^{l,\nu})_{33}\end{array}\right]
\end{eqnarray} 
and
\begin{eqnarray}
 \mathbf{N_{1}}=(A^{\Phi})^{T}\otimes (A^{L})^{\dag}\otimes (A^{l})^{T},\\\mathbf{N_{2}}=(A^{\Phi})^{\dag}\otimes (A^{L})^{\dag}\otimes (A^{\nu})^{T}.
\end{eqnarray}

Thus in order to get the desired form of Yukawa couplings one should simply find the eigensubspaces of matrices $\mathbf{N_{1}}$ and $\mathbf{N_{2}}$ corresponding to the eigenvalue 1.

However, the group has got many elements, each of them has got its own set of irreducible representations. Thus, should we solve Eq.(4) for each group's element  separately? 
Luckily, the answer to this question is negative. It turns out that one has to take only  generators' representations in order to obtain the Yukawa couplings which are invariant with respect to the whole  group $G_{F}$ \cite{ludl2}.
\begin{thh} 
If the relation :
$$
\sum_{i=1,2} (A^{\phi})_{i,k}(A^{L})^{\dag}_{\alpha\gamma}(h^{l}_{i})_{\gamma\delta}(A^{l})_{\delta\beta}=(h^l_{k})_{\alpha\beta}
$$
 holds 
for the generators' representations of some group $G_{F}$ then it holds for the  representations of all group's elements.
\end{thh}
\textsl{Proof}: Every element of the group $G$ can be presented as  a unique combination of its generators \{A,B,C,..\} :\\
$$
G=A^{n_{1}}B^{n_{2}}C^{n_{3}}...
$$
where $n_{1},n_{2},n_{3},...$ are  natural numbers. \\
Therefore it is sufficient to prove that if the relations :
$$
\sum_{i=1,2} (A^{\Phi})_{ik}(B)(A^{L})^{\dag}_{\alpha\gamma}(B)(h^{l}_{i})_{\gamma\delta}(A^{l})_{\delta\beta}(B)=(h^l_{k})_{\alpha\beta},
$$
$$
\sum_{i=1,2} (A^{\Phi})_{ik}(C)(A^{L})^{\dag}_{\alpha\gamma}(C)(h^{l}_{i})_{\gamma\delta}(A^{l})_{\delta\beta}(C)=(h^l_{k})_{\alpha\beta}
$$
are fulfilled for the representations $A^{\Phi},A^{L},A^{R}$ of some generators $B,C$ then they are also valid for the representations $A^{\Phi},A^{L},A^{R}$ of their product:
$$
\sum_{i=1,2} (A^{\Phi})_{ik}(BC)(A^{L})^{\dag}_{\alpha\gamma}(BC)(h^{l}_{i})_{\gamma\delta}(A^{l})_{\delta\beta}(BC)=(h^{l}_{k})_{\alpha\beta}.
$$
The above-mentioned statement holds indeed:
\begin{multline}
\sum_{i=1,2} (A^{\Phi})_{ik}(BC)(A^{L})^{\dag}(BC)(h^{l}_{i})(A^{l})(BC)=\\
=\sum_{i,m=1,2} (A^{\Phi})_{ik}(B)(A^{\Phi})_{km}(C)(A^{L})^{\dag}(C)(A^{L})^{\dag}(B)
(h^{l}_{i})(A^{l})(B)(A^{l})(C)=\\=\sum_{i,m=1,2}(A^{\Phi})_{km}(C)(A^{L})^{\dag}(C)
(A^{\Phi})_{ik}(B)(A^{L})^{\dag}(B)(h^{l}_{i})(A^{l})(B) (A^{l})(C)
=\\=\sum_{i,m=1,2} (A^{\Phi})_{km}(C)(A^{L})^{\dag}(C)(h^{l}_{k})(A^{l})(C)=(h^{l}_{m})
\end{multline}
We arrive thereby to the conclusion that only group's generators are necessary to find invariant Yukawa couplings $h_{1,2}^{l,\nu}$. After solving  Eq.(4) for each of the generators one will get some set of eigenspaces $\mathcal{W}_{i}$ to the eigenvalue 1 (this eigenvalue is  in general degenerate). In order to get the appropriate solution, one has to find the common eigenspace $\mathcal{U}$ of these individual eigenspaces  . Then, it is necessary to find the basis vector of $\mathcal{U}$ which constitutes the ultimate result.

At this point, knowledge about the form of Yukawa matrices makes the construction of mass matrices feasible: 
\begin{eqnarray}
M^{l}_{\alpha,\beta}=\frac{1}{\sqrt{2}}(v_{1}h_{1}^{l}+v_{2}h_{2}^{l}),\nonumber\\
M^{\nu}_{\alpha\beta}=\frac{1}{\sqrt{2}}(v_{1}h_{1}^{\nu}+v_{2}h_{2}^{\nu}).
\end{eqnarray}
Then, we are able to find the $U_{PMNS}$ mixing matrix, which is composed of the matrices  that diagonalize $M^{l}$ and $M^{\nu}$:
\begin{gather}
V_{L}^{l\dag}M^{l}V^{l}_{R}=diag(m_{e}, m_{\mu}, m_{\tau}),\nonumber\\
V_{L}^{\nu\dag}M^{\nu}V^{\nu}_{R}=diag(m_{\nu_{e}}, m_{\nu_{\mu}}, m_{\nu_{\tau}}),\nonumber\\
U_{PMNS}=V_{L}^{l\dag}V_{L}^{\nu}.
\end{gather}
\section{The search for desired groups}
It is clear, that in order to get the form of Yukawa matrices one needs to find the irreducible representations of the groups  which meet our requirements. 

First of all, we want our group to be discrete, non-abelian and to  be a subgroup of U(3) (similarly to $A_{4}$).
Then, it is quite  obvious that the group  should possess 2 and 3 dimensional irreducible representations, which are necessary for construction of  Eq.(3).

To find such a group one can make use of $GAP$: the programme for discrete algebra computation available on the website \cite{GAP}. One would also need the $Small Group$ library \cite{small2} and $REPSN$ package \cite{repsn} which serve as indispensable tools for our purposes.

In order to perform the calculation it is recommended to benefit from some mathematical theorems:
\begin{thh}{}
The order of the finite group $\mathcal{G}$ is divisible by dimension of its irreducible representation.
  \end{thh} 
\begin{thh}
A finite group $\mathcal{G}$ is isomorphic to a finite subgroup of U(3) if and only if it possesses a faithful (irreducible or reducible) 3-dimensional representation.
\end{thh}
Since our group has got 2 and 3-dimensional irreducible representations, following the second theorem, one should first verify that the group's order is divided by 6.

The proof of the third theorem can be found in many textbooks about finite groups (see, for example, \cite{grimus}). On the other hand, the justification of the last statement is quite intuitive. In case of finite groups, every representation has got its unitary equivalent (it is possible to present the matrices of the given representation in such a basis to make them unitary). Therefore, the faithfulness assures that every element $g$ of $G$ can uniquely be expressed as a unitary matrix (see \cite{small} for a more profound analysis of this theorem).

\section{The search for Yukawa matrices}

After finding the irreducible representations of $G_{F}$, one has got almost everything to get to know how Yukawa matrices look like.
Since the equation restricting the form of Yukawa matrices  has the form of eigenproblems to the eigenvalue 1, its solution is rather straightforward. One can easily compute the eigenspace $\mathcal{W}_{i}$ for
each of the generators. The problem arises, when one wants to deal with the common eigenspace $\mathcal{U}$ of the eigenspaces $\mathcal{W}_{i}$. In order to perform this calculation, one can follow very simple
algorithm (it is not necessarily the most efficient one). To simplify the considerations, let us assume that one has got two n-dimensional spaces $\mathcal{S}$ and $\mathcal{T}$. Each space is spanned by its basis vectors:
$\{s_{1},s_{2},...,s_{n}\}$ and 
$\{t_{1},t_{2},..,t_{n}\}$  respectively. Vector $\vec{a}$ being the part of the common subspace $\mathcal{P}=\mathcal{S}\cap \mathcal{T}$ can be equivalently expressed
in  two basis:
\begin{eqnarray}
\vec{a}=a_{1}\vec{s_{1}}+a_{2}\vec{s_{2}}+...+a_{n}\vec{s_{n}},\nonumber\\
\vec{a}=a_{1}'\vec{t_{1}}+a_{2}'\vec{t_{2}}+...+a_{n}'\vec{t_{n}},
\end{eqnarray}
where $a_{1},a_{2},...,a_{n},a'_{1},a_{2}',...,a_{n}'$ are some complex coefficients.

Denoting the basis vectors of space $\mathcal{S}$ via their components as:
$$
\vec{s_{1}}=\left[\begin{array}{c}s_{11}\\s_{12}\\.\\.\\.\\s_{1n}\end{array}\right],\vec{s_{2}}=\left[\begin{array}{c}s_{21}\\s_{22}\\.\\.\\.\\s_{2n}\end{array}\right],\dots,\vec{s_{n}}=\left[\begin{array}{c}s_{n1}\\s_{n2}\\.\\.\\.\\s_{nn}\end{array}\right]
$$
and similarly for $\mathcal{T}$, one can rewrite the equality following from Eq.(11) as a set of equations:
\begin{eqnarray}
a_{1}s_{11}+a_{2}s_{21}+...+a_{n}s_{n1}-a_{1}'t_{11}-a_{2}'t_{21}-...-a_{n}'t_{n1}=0,\nonumber\\
a_{1}s_{12}+a_{2}s_{22}+...+a_{n}s_{n2}-a_{1}'t_{12}-a_{2}'t_{22}-...-a_{n}'t_{n2}=0,\nonumber\\
.\qquad.\qquad.\qquad.\qquad.\qquad.\qquad.\qquad.\qquad.\qquad.\qquad.\qquad.\nonumber\\
.\qquad.\qquad.\qquad.\qquad.\qquad.\qquad.\qquad.\qquad.\qquad.\qquad.\qquad.\nonumber\\
.\qquad.\qquad.\qquad.\qquad.\qquad.\qquad.\qquad.\qquad.\qquad.\qquad.\qquad.\nonumber\\
a_{1}s_{1n}+a_{2}s_{2n}+...+a_{n}s_{nn}-a_{1}'t_{1n}-a_{2}'t_{2n}-...-a_{n}'t_{nn}=0,
\end{eqnarray}
which  can be readily presented in the matrix form:
$$
\left[\begin{array}{cccccccc}
s_{11}&s_{21}&...&s_{n1}&-t_{11}&t_{21}&...&t_{n1}\\
s_{12}&s_{22}&...&s_{n2}&-t_{12}&t_{22}&...&t_{n2}\\
...&...&...&...&...&...&...&...\\
s_{1n}&s_{2n}&...&s_{nn}&-t_{1n}&t_{2n}&...&t_{nn}
\end{array}\right]
\left[\begin{array}{c}
a_{1}\\
a_{2}\\
.\\
.\\
.\\
a_{n}\\
a_{1}'\\
a_{2}'\\
.\\
.\\
.\\
a_{n}'
\end{array}\right]=\left[\begin{array}{c}
0\\
0\\
.\\
.\\
.\\
0\\
0\\
0\\
.\\
.\\
.\\
0
\end{array}\right]
$$
Therefore, in order to find the common eigenspace one needs to calculate the coefficients $a_{1},a_{2},...,a_{n},a_{1}',...,a_{2}'$ (one has to 
find the null space of the matrix composed of these coefficients).

The generalization of this algorithm for  the case of several spaces is trivial. One needs to repeat the described steps iteratively: one should  take 2 spaces at first, find the null space, take another, calculate a new null space, etc.
\section{Interpretation of Yukawa matrices}

It turns out, that the Yukawa matrices calculated according to Eq.(3) have got simple mathematical interpretation \cite{ludl2}. The following theorem clearly  illustrates this statement:
\begin{thh}
If the coefficients $h$ constitute the solution to the equation:
$$
(D^{*}\otimes A\otimes B)h=h
$$
then they can interpreted as the Clebsch-Gordan coefficients for the decomposition:
$$
A\otimes B=\oplus_{D} D
$$
where $A,B$ are 3-dimensional irreducible representation  and one of $D$ is identified with 2-dimensional irreducible representation.
\end{thh}
\textsl{Proof}\\
To start with, let us define the orthonormal  basis for appropriate matrix representations:
\begin{itemize}
\item $\{e^{A}_{k}\}$ for representation $A$, k=1,2,...,$N_{A}$ (spanning the vector space $\mathcal{V_{A}}$),
\item  $\{e^{B}_{m}\}$ for representation $B$, k=1,2,...,$N_{B}$ (spanning the vector space $\mathcal{V_{B}}$),
\item $\{e^{D}_{n}\}$ for representation $D$, n=1,2,...,$N_{A}N_{B}$ (spanning the vector space $\mathcal{V_{A}}\otimes \mathcal{V_{B}})$,
\item $\{e^{AB}_{km}=e^{A}_{k}\otimes e^{B}_{m}\}$ for representation $A\otimes B$ (spanning the vector space $\mathcal{V_{A}}\otimes \mathcal{V_{B}}$).
\end{itemize}
Since the vector basis $\{e^{D}\}$ and $\{e^{AB}\}$ act in the same space $(\mathcal{V_{A}}\otimes \mathcal{V_{B}})$ one can write:
\begin{equation}
e^{AB}_{km}=\sum_{D,n}\alpha^{D}_{n}e^{D}_{n},\quad\quad e^{D}_{n}=\sum_{m,n}\beta^{AB}_{mn}e^{AB}_{mn}.
\end{equation}
At this moment, it is convenient to define the Clebsch-Gordan coefficients $C^{DAB}_{ijk}$ as:
\begin{equation}
C^{DAB}_{ijk}=(\alpha^{D}_{i})^{*}=\beta^{AB}_{jk}=(e_{jk}^{AB})^{\dag}e^{D}_{i}.
\end{equation}
From Eq.(12) and Eq.(13) it is possible to investigate the form of group's elements in the $\{e^{LR}_{km}\}$ basis:
\begin{gather}
(e^{AB}_{ij})^{\dag}(A\otimes B)e^{AB}_{kl}=((e^{A})^{\dag}_{i}\otimes (e^{B}_{j})^{\dag})(A\otimes B)(e^{A}_{k}\otimes e^{B}_{l})=(e^{A})^{\dag}_{i}Ae^{A}_{k}\otimes (e^{B})^{\dag}_{j}Be^{B}_{j}=A_{ik}B_{jl}\nonumber\\
(e^{AB}_{ij})^{\dag}(\oplus_{D} D)e^{AB}_{kl}=\sum_{D',D'',m',m''} (\alpha^{D'}_{m'})^{*}(\alpha^{D''}_{m''})(e^{D'}_{m'})^{\dag}(\oplus_{D}D)e^{D''}_{m''}=\sum_{D',D'',m',m''}(\alpha^{D'}_{m'})^{*}\alpha^{D''}_{m''}\delta_{D',D}\delta_{D'',D}\nonumber\\
\times ((e^{D}_{m'})^{\dag}De^{D}_{m''})=\sum_{D,m',m''} (\alpha^{D}_{m'})^{*}\alpha^{D}_{m''}D_{m',m''}=\sum_{D,m',m''}C^{DAB}_{m'ij}(C^{DAB}_{m''kl})^{*} D_{m',m''}.\nonumber
\end{gather}
The above calculations lead us to very important relation:
\begin{equation}
A_{ik}B_{jl}=(A\otimes B)_{ij,kl}=\sum_{D',m',m''}C^{D'AB}_{m'ij}(C^{D'AB}_{m''kl})^{*}D'_{m',m''}.
\end{equation}
Multiplying both sides of Eq.(15) by $\sum_{k,l}C^{DAB}_{mkl}$ and making use of the orthogonality relation for Clebsch-Gordan coefficients ($\sum_{i,j} (C^{DAB}_{mij})^{*}C^{D'AB}_{m'ij}=\delta_{D,D'}\delta_{m,m'}$) one arrives to:
\begin{equation}
(\sum_{k,l}(A\otimes B)_{ij,kl} C^{DAB}_{mkl})=\sum_{D',m',m''}C^{D'AB}_{m'ij}D'_{m',m''}\delta_{D,D'}\delta_{m,m''}=\sum_{m'} C^{DAB}_{m'ij}D_{m',m}.
\end{equation}
After multiplication of the previous equation by $(\sum_{m}D^{\dag}_{mn})$ one gets:
\begin{equation}
\sum_{mkl}(D_{n,m})^{*}(A \otimes B)_{ij,kl}C^{DAB}_{mkl}=C^{DAB}_{nij},
\end{equation}
which is equivalent to:
\begin{equation}
\sum_{mkl}(D^{*}\otimes A\otimes B)_{nij,mkl}C^{DAB}_{mkl}=C^{DAB}_{nij}\Leftrightarrow \sum_{mkl}(D^{T}\otimes (A)^{\dag}\otimes (B)^{\dag})_{nij,mkl}C^{DAB}_{mkl}=C^{DAB}_{nij}.
\end{equation}
Denoting the representation $D$ by $A^{\Phi}$, replacing $B\rightarrow (A^{R})^{*}$ and $A\rightarrow A^{L}$ one finds that previous equation is completely analogical to the first part of Eq.(4). On the other hand, when it comes to the second part of Eq.(4) it is easy to notice that we get the equivalence if we assume  $D=(A^{\Phi})^{*}$.

This theorem indicates that in order to find out if any solution for Eq.(4) exists, all we need to do, is the investigation of the Clebsch-Gordan decompositions:
\begin{eqnarray}
A^{L}\otimes A^{l}=\oplus_{D} D,\\
A^{L}\otimes A^{\nu}=\oplus_{D} D.
\end{eqnarray}
Therefore, if one finds the representations $A^{L}$ and $A^{l}$,  direct product of which gives $A^{\Phi}$ (one of the 2-dimensional representations among $D$) in the decomposition, then we have the guarantee that some solution to Eq.(4) exists .The similar situation takes place, when it comes to $A^{L}\otimes A^{\nu}$. The only exception lies in the fact, that we need to look for $(A^{\Phi})^{*}$ in the decomposition. All these theorems  can be easily verified by $GAP$.
\section{Conclusions}
To sum up, it is necessary to replace $TBM$ mixing, which according to the experiments carried out in 2012, is no longer valid. Basing on the literature, we have developed the tools which are indispensable to search for the answer
in the models with more Higgs doublets (these methods can be trivially generalized into the case of more Higgs doublets). Our results will be presented in the forthcoming paper. In the meantime, the preliminary outcomes can be found in \cite{richter}.

This work has been supported by the Polish Ministry of Science and Higher Education under grant No. UMO-2013/09/B/ST2/03382.

\end{document}